\documentclass[10pt,aps,prx,twocolumn,superscriptaddress,floatfix,showpacs,longbibliography]{revtex4-1}

\usepackage{graphicx}
\usepackage{amsfonts}
\usepackage{amssymb}
\usepackage{amsmath}
\usepackage{txfonts}
\usepackage{lipsum}
\usepackage{color}
\usepackage{wasysym}
\usepackage{hyperref}
\usepackage{bbold}
\usepackage{etoolbox} 
\usepackage[normalem]{ulem}
\usepackage{mathtools}
\usepackage{multirow}
\usepackage{hhline}
\usepackage{mathrsfs} 
\usepackage{soul}

\usepackage{letltxmacro}
\LetLtxMacro{\oldsqrt}{\sqrt}
\renewcommand{\sqrt}[2][\mkern8mu]{\mkern-6mu\mathop{}\oldsqrt[#1]{#2}}

\begin{document}
\title{
Nonlinear optical study of commensurability effects in graphene-hBN heterostructures
}

\author{E. A. Stepanov}
\thanks{These authors contributed equally}
\affiliation{I. Institute of Theoretical Physics, Department of Physics, University of Hamburg, Jungiusstrasse 9, 20355 Hamburg, Germany}
\affiliation{Theoretical Physics and Applied Mathematics Department, Ural Federal University, Mira Street 19, 620002 Ekaterinburg, Russia}

\author{S. V. Semin}
\thanks{These authors contributed equally}
\affiliation{Radboud University, Institute for Molecules and Materials, 6525AJ Nijmegen, The Netherlands}

\author{C. R. Woods}
\affiliation{School of Physics and Astronomy, University of Manchester, Oxford Road, Manchester, M13 9PL, UK}
\affiliation{National Graphene Institute, University of Manchester, Oxford Road, Manchester, M13 9PL, UK}

\author{M. Vandelli}
\affiliation{I. Institute of Theoretical Physics, Department of Physics, University of Hamburg, Jungiusstrasse 9, 20355 Hamburg, Germany}
\affiliation{The Hamburg Centre for Ultrafast Imaging, Luruper Chaussee 149, 22761 Hamburg, Germany}
\affiliation{Max Planck Institute for the Structure and Dynamics of Matter, Center for Free Electron Laser Science, 22761 Hamburg, Germany}

\author{A. V. Kimel}
\affiliation{Radboud University, Institute for Molecules and Materials, 6525AJ Nijmegen, The Netherlands}

\author{K. S. Novoselov}
\affiliation{School of Physics and Astronomy, University of Manchester, Oxford Road, Manchester, M13 9PL, UK}
\affiliation{National Graphene Institute, University of Manchester, Oxford Road, Manchester, M13 9PL, UK}
\affiliation{Chongqing 2D Materials Institute, Liangjiang New Area, Chongqing, 400714, China}

\author{M. I. Katsnelson}
\affiliation{Radboud University, Institute for Molecules and Materials, 6525AJ Nijmegen, The Netherlands}
\affiliation{Theoretical Physics and Applied Mathematics Department, Ural Federal University, Mira Street 19, 620002 Ekaterinburg, Russia}

\begin{abstract}
Second-order nonlinear optical response allows to detect different properties of the system associated with the inversion symmetry breaking. Here, we use a second harmonic generation effect to investigate the alignment of a graphene/hexagonal Boron Nitride heterostructure. To achieve that, we activate a commensurate-incommensurate phase transition by a thermal annealing of the sample. We find that this structural change in the system can be directly observed through a strong modification of a nonlinear optical signal. This result reveals the potential of a second harmonic generation technique for probing structural properties of layered systems.
\end{abstract}

\maketitle

Two-dimensional (2D) materials attract a lot of attention due to their remarkable characteristics. These systems have a rich variety of structural modifications and chemical compositions, which results in a high tunability of their physical properties.
Moreover, collective many-body effects that are essentially strong in two dimensions give rise to nontrivial phases of matter, which appear promising for various electronic and opto-electronic applications. 
For example, transition metal dichalcogenides (TMDs) reveal competing charge~\cite{PhysRevB.82.075130, C5CP01326G} and spin-ordered~\cite{PhysRevB.93.054429, PhysRevB.94.035120}, as well as Mott insulating~\cite{cho2016nanoscale, ma2016metallic} states in the mono- and multilayered phases. Also, 2D systems show a superconducting behaviour~\cite{Uchihashi_2016} that was theoretically predicted for black phosphorus~\cite{Shao_2014, Ge_2015} and antimony~\cite{PhysRevB.99.064513}, and experimentally observed in black phosphorus~\cite{zhang2017intercalant} and various TMDs~\cite{Ye1193, Lu1353, xi2016ising, Lu3551}.

In a multilayered phase, characteristics of quasi-2D systems depend not only on internal properties of individual layers. 
Here, the number of layers and, what is even more important, the form of their stacking also plays a significant role. 
A famous example is a twisted bilayer graphene that at a magic misorientation angle of $1.1^{\circ}$ between two graphene lattices drastically changes electronic properties and becomes superconducting. 
This remarkable effect has been first theoretically predicted~\cite{Bistritzer12233, PhysRevB.82.121407} and recently confirmed experimentally~\cite{cao2018unconventional, cao2018correlated, Yankowitz1059}. Also, a strong correlation between the interlayer stacking and optical properties has been reported for MoS$_2$~\cite{hsu2014second, Yin488} and hexagonal Boron Nitride (hBN)~\cite{kim2013stacking, yang2019stacking}.

Recently, it became possible to experimentally produce hybrid systems combining different two-dimensional materials in a layered heterostructure~\cite{2dmat}. 
This gives even more freedom in designing materials with special characteristics in a controlled way.
For heterostructures, the effect of different stacking is much less explored than in the case of multilayered homostructures. However, this issue should also be of a crucial importance as it involves different types of materials. For instance, a change of a misorientation angle between graphene and hBN lattices results in a commensurate-incommensurate phase transition, which strongly affects a crystal symmetry and electronic properties of the system~\cite{woods2014commensurate}.
Therefore, it is very important to have a simple yet sensitive tool that provides the information about the structural properties of a material, because even a small mismatch of different layers dramatically changes characteristics of the system.

\begin{figure}[t!]
\includegraphics[width=0.95\linewidth]{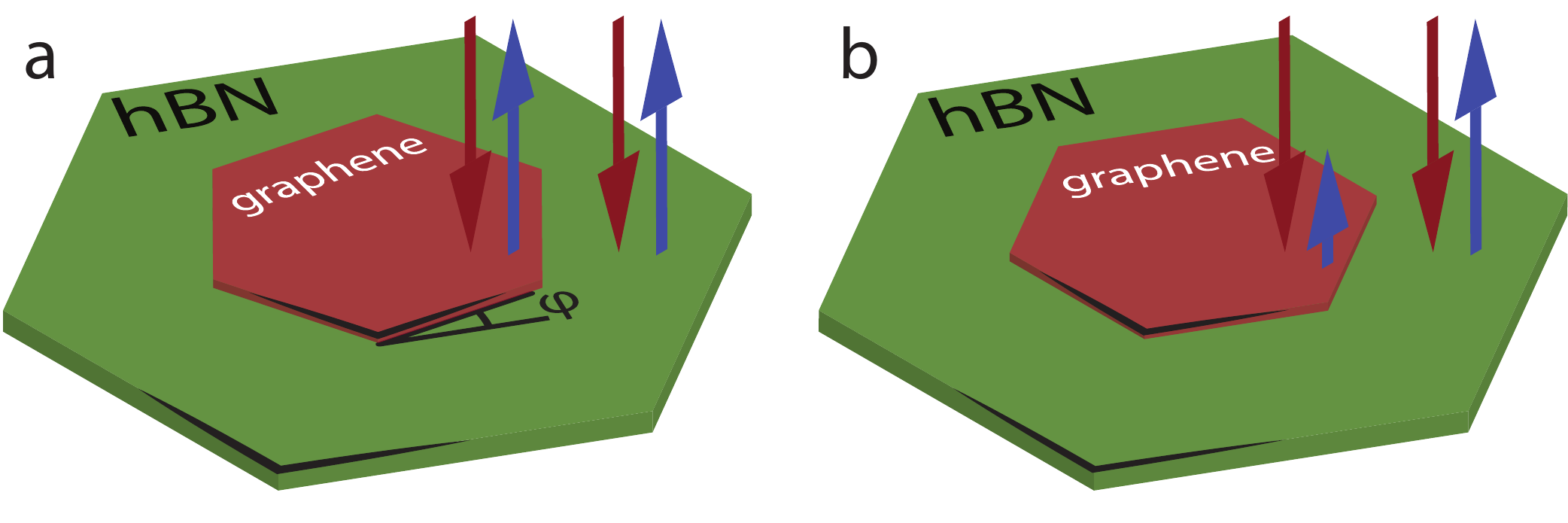}
\caption{\label{fig:Scheme}
Sketch of the experiment. Green and red hexagonal tiles represent hBN and graphene, respectively. Red arrows depict the incident 800\,nm light. Blue arrows indicate the collected SHG response at 400\,nm from different parts of the sample. In the incommensurate phase (a), the signal of the SHG is uniform for the entire sample. The strong modification of the SHG response is expected after the structural phase transition (b) from the aligned graphene area.
}
\end{figure}

Conventional experimental techniques that allow to probe various electronic and symmetry properties of nanostructures are Raman spectroscopy~\cite{woods2014commensurate} and transport measurements~\cite{Gorbachev448}. 
Raman technique is by no means a powerful method that, however, requires an accurate interpretation of results, which might not be straightforward even for a simple case of graphene~\cite{Malard2009, FERRARI200747}. Also, Raman signals are integrated over rather thick layer of material ($>$500\,nm) which may result in additional difficulties during data analysis, as relevant signals might originate from buried layers or interfaces~\cite{Malard2009}.
Transport experiments are much more difficult to perform as they require additional sample preparationtion steps and corresponding facilities.
In this regard, an optical second harmonic generation (SHG) is a very promising tool for investigation of different structural properties of 2D materials, being sensitive to the inversion symmetry breaking in the system and, therefore, crucially depends on the number of layers, stacking, alignment and etc.
Thus, the SHG has already been used for description of non-centrosymmetric 2D systems, such as MoS$_2$~\cite{li2013probing, hsu2014second, ultrastrong, PhysRevB.87.201401, PhysRevB.87.161403, Yin488} and WS$_2$~\cite{janisch2014extraordinary, seyler2015electrical, expTMDs}. Also, the optical SHG response has been observed from graphene, where the inversion symmetry was broken by a presence of an electric field~\cite{PhysRevB.85.121413, PhysRevB.91.205405}. 

Experimental SHG studies of quasi-2D homostructures are almost entirely dedicated to mono-, or few-layered systems. An exception is a very recent study where a nonzero SHG response from rather thick hBN (ca. hundred layers) flake has been reported~\cite{2019arXiv190209060K}. 
SHG from layered heterostructures is much less studied, both theoretically~\cite{marg, PhysRevB.99.165432} and experimentally~\cite{doi:10.1063/1.3275740, PhysRevB.82.125411, an2013enhanced}, and is mostly focused on graphene/graphite films, where the inversion symmetry is broken by the interface with a substrate. 

An effect of the stacking on the nonlinear optical response in layered heterostructures has not been investigated experimentally so far.
Here, we address this important question for the cases of mono- and multilayer graphene disposed on an insulating hBN substrate. 
We show that the change of the SHG response clearly indicates the commensurate-incommensurate phase transition for both considered heterostructures. The change of the alignment in the case of the monolayer graphene has been additionally confirmed by Raman measurements. At the same time, Raman spectroscopy was not able to characterize the structural change in the multilayer graphene/hBN heterostructure. This result suggest that the SHG can be used as a simple and efficient method for probing structural properties and alignment of layered heterostructures.

\begin{figure}[t!]
\includegraphics[width=0.95\linewidth]{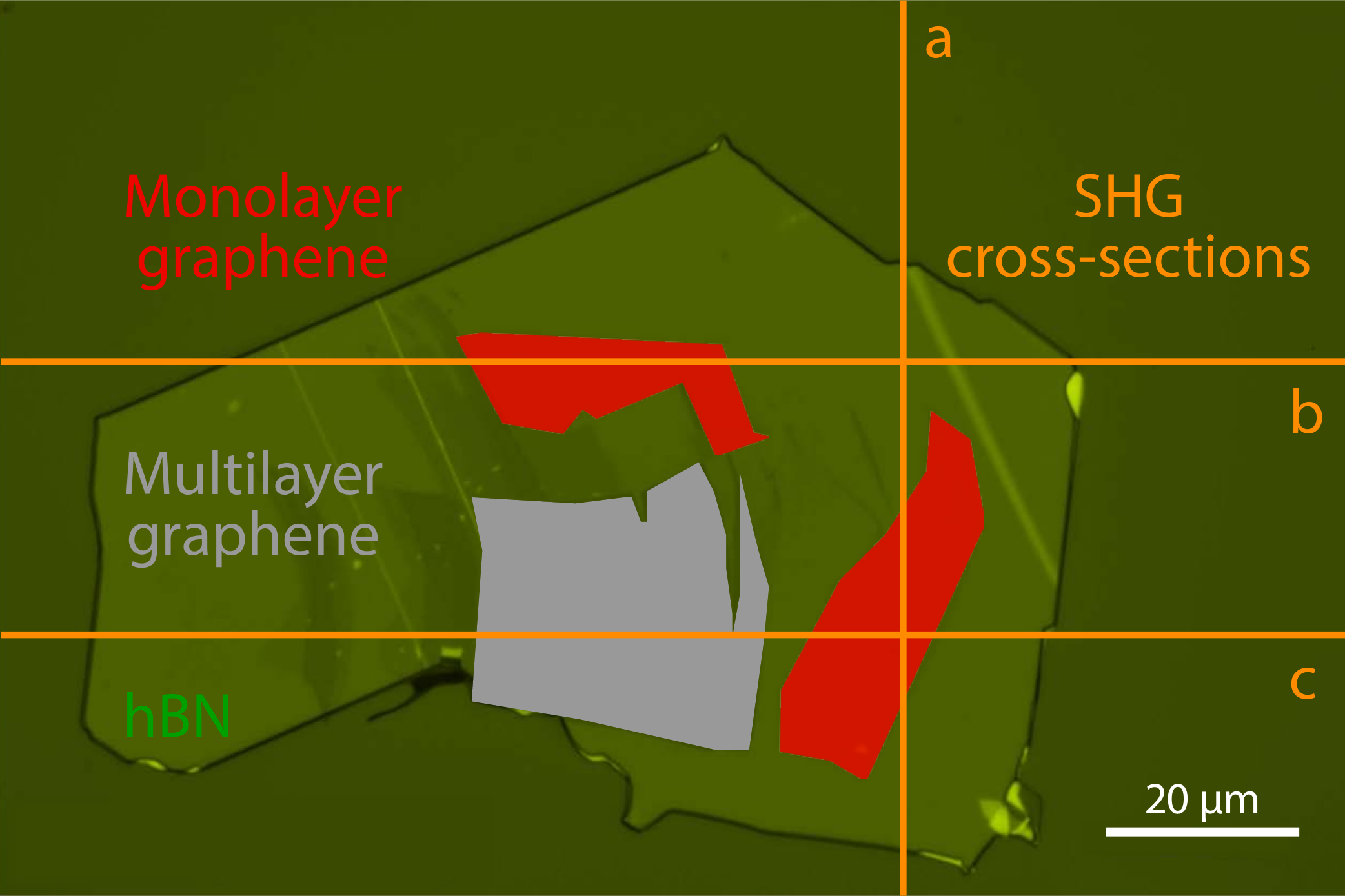}
\caption{\label{fig:Sample}
Optical image of the graphene/hBN sample. 
We use a green filter to enhance contrast. Monolayer graphene regions are shaded red. Gray square area corresponds to a multilayer graphene. The substrate hBN crystal is outlined in light green. Orange lines indicate cross sections shown in Fig.~\ref{fig:cut}. Scale bar is 20\,um.}
\end{figure}

We investigate a nonlinear optical response from graphene flakes placed on top of a hexagonal Boron Nitride (hBN) substrate.
In general, a small mismatch between lattice constants of these two materials leads to hexagonal moir\'e patterns formed in the system~\cite{xue2011scanning, yankowitz2012emergence}. 
If a misorientation angle between graphene and hBN lattices is less than $1^{\circ}$, a commensurate stacking is energetically more preferable~\cite{woods2014commensurate}. In this case, electronic properties of the system can be effectively described by Dirac electrons with a nonzero mass~\cite{PhysRevB.87.245408, PhysRevB.84.195414, PhysRevLett.115.186801}. The latter appears as a result of an inversion symmetry breaking in a graphene/hBN heterostructure. Importantly, the value of the mass varies in space with a period of moir\'e pattern, but the average value of the mass stays nonzero~\cite{PhysRevB.89.201404, PhysRevB.84.195414, PhysRevLett.115.186801}. Apart from the theoretical prediction, the nonzero average mass in the commensurate phase was also observed in transport experiments~\cite{Gorbachev448}. At larger values of the misorientation angle the system undergoes a structural phase transition towards an incommensurate phase~\cite{woods2014commensurate}. In this phase, the inversion symmetry is still broken locally, but the average value of the mass becomes zero~\cite{PhysRevB.89.201404}.
Since a typical focused laser spot is much larger than an interatomic distance, only the average value of the effective mass can be probed in optical experiments. The SHG is known to be very sensitive to the lack of inversion symmetry in the point group of the medium. Therefore, such a nonlinear optical technique a very promising tool for detection of the alignment of such layered heterostructures.

\begin{figure}[t!]
\includegraphics[width=0.95\linewidth]{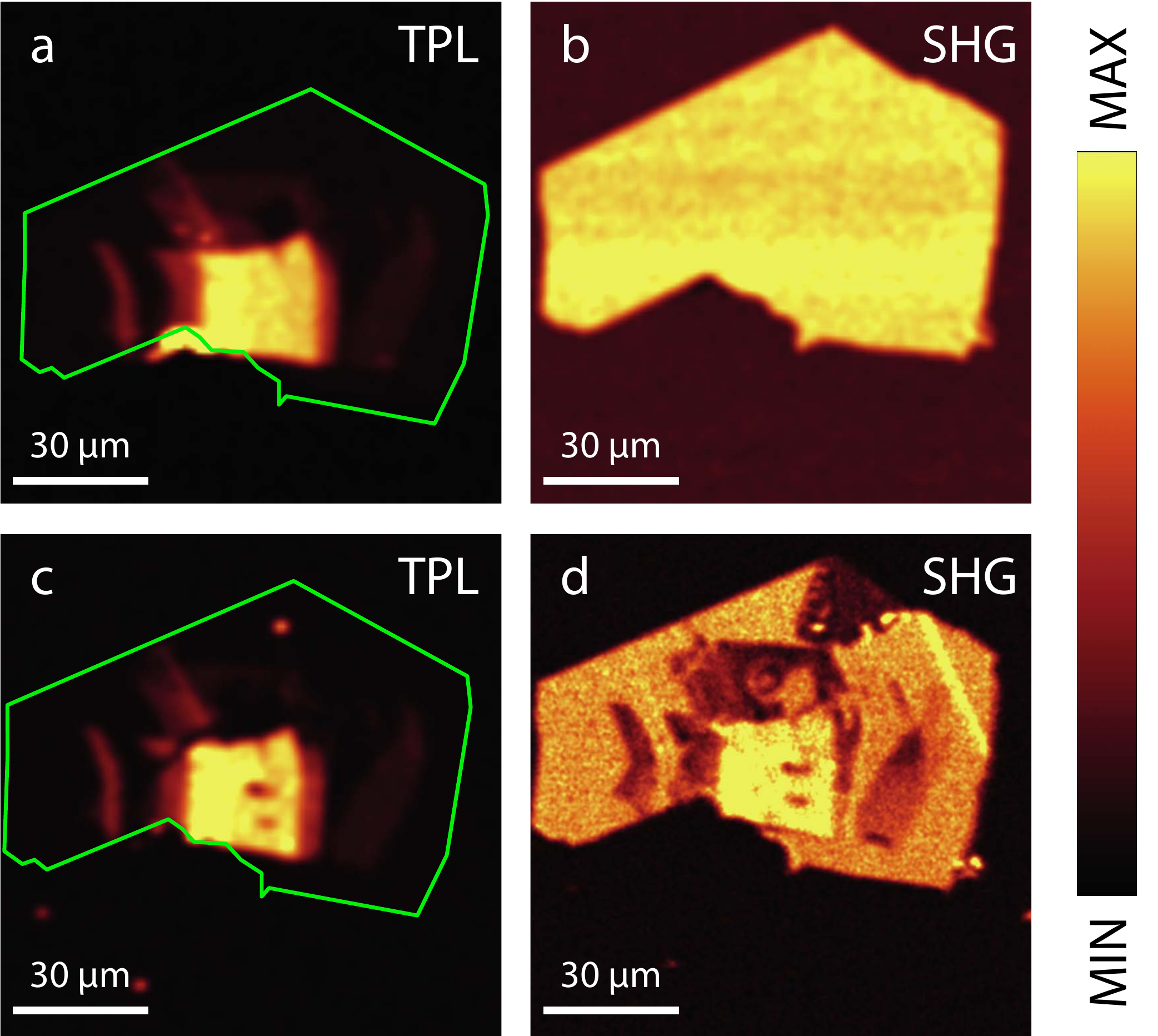}
\caption{
TPL (a, c) and SHG (b, d) signals from incommensurate (a, b) and commensurate (c, d) stacking of graphene/hBN heterostructure. Scans were performed on the same sample before and after the structural phase transition. Color bar depicts the intensity of the nonlinear response in arbitrary units. A lighter color indicates a larger value of a signal. Green line outlines the hBN sample area.}
\label{fig:2x2}
\end{figure}

Schematic representation of the experiment is shown in Fig.~\ref{fig:Scheme}. The optical image of the considered sample obtained from optical microscope through a $\times100$ objective is shown in Fig.~\ref{fig:Sample}. Here, red areas highlight single-layer graphene flakes placed on top of the hBN substrate depicted by a light green color. The incommensurate phase for both flakes is confirmed by Raman spectroscopy through the broadening of the 2D-peak in the Raman spectrum~\cite{eckmann2013raman}. A central gray square area corresponds to a multilayer graphene, which alignment with respect to the hBN was not possible to define by Raman measurements.
Details of the sample preparation and Raman experiments can be found in the Section ``Methods''. 
A combined optical response has been measured from the sample excited by a 800 nm femtosecond laser. A two-photon luminescence (TPL) signal that was collected at $390-650$ nm is shown in Fig.~\ref{fig:2x2}\,a. A narrow bandpass filter centered at 400 nm ($\pm$ 20 nm) was used to detect the SHG signal only (see Fig.~\ref{fig:2x2}\,b). Description of the experimental set-up can also be found in the Section ``Methods''.

Since the TPL is an incoherent process, it is not sensitive to the point group of the medium. Thus, we observe the TPL response only from graphene areas. Here, the strongest signal comes from a multilayer graphene as it has more complex band structure than its single-layer realization. The hBN does not contribute to the TPL, because the corresponding excitation energy $\hbar\omega\sim 1.9-3.2$\,eV is lower than the band gap in this material, which is about 6\,eV~\cite{PhysRevB.51.6868, cassabois2016hexagonal}. Thus, a nonzero TPL signal only confirms the presence of graphene flakes on the hBN substrate. As expected, this incoherent optical process is not sensitive to the alignment of the considered heterostructure.

On contrary, the SHG is a coherent process, whose efficiency must obey the symmetry principle~\cite{nye1985physical}. Hence SHG can serve as a probe of symmetries of the point group in the studied medium. For considered materials, the latter is mediated by the value of the average mass (band gap)~\cite{PhysRevB.99.165432}. We observe that the intensity of the SHG response shown in Fig.~\ref{fig:2x2}\,b is uniform for the entire sample. This fact suggests that there is no inversion symmetry breaking associated with the interaction of graphene with the hBN substrate. 
Thus, we find that single-layer graphene flakes are not aligned with the hBN, which is consistent with Raman measurements. 
This result also confirms that the average mass of graphene, which is effectively probed by the SHG, is zero in the incommensurate phase. Importantly, we also do not observe any change of the SHG signal from the multilayer graphene area. This indicates that the multilayer graphene flake is not aligned with hBN substrate as well. 
As a consequence, the SHG response from the sample is related only to the hBN.

We note that the SHG from a thick hBN is strong. This can be explained by a simple phenomenological model. The SHG arises from the nonlinear polarization ${\bf P}(2\omega)$ induced by an incident laser field ${\bf E}(\omega)$. To the leading orders in ${\bf E}(\omega)$, this polarization can be written as~\cite{HEINZ1991353}:
\begin{align*}
P_{i}(2\omega) = \chi^{d}_{ijk}E_{j}(\omega)E_{k}(\omega) + \chi^{q}_{ijkl}E_{j}(\omega)\nabla_{k}E_{l}(\omega),
\end{align*}
where the first and second terms describe the electric-dipole and quadrupole contributions to the SHG, respectively. The total SHG response from the system can be divided into the surface and bulk parts. 
According to Neumann's principle~\cite{nye1985physical}, the electron-dipole SHG process is allowed only for a non-centrosymmetric medium. This results in a strong second-order optical response from the surface of hBN~\cite{PhysRevB.99.165432}. On contrary, the bulk of hBN obeys the inversion symmetry and contributes to the SHG only through the quadrupole polarization. 
Although the latter is much smaller than the electric-dipole one, in birefringent materials the total quadrupole contribution from the bulk can be of the same order of magnitude as the surface SHG due to the phase-matching~\cite{PhysRevLett.51.1983, HEINZ1991353}. 
Moreover, in some cases these two processes can hardly be separated from each other~\cite{PhysRevB.38.7985, HEINZ1991353}.
However, the strength and the relative phase of the quadrupole contribution to the SHG strongly depends on the particular structure of the bulk.

\begin{figure}[t!]
\includegraphics[width=0.9\linewidth]{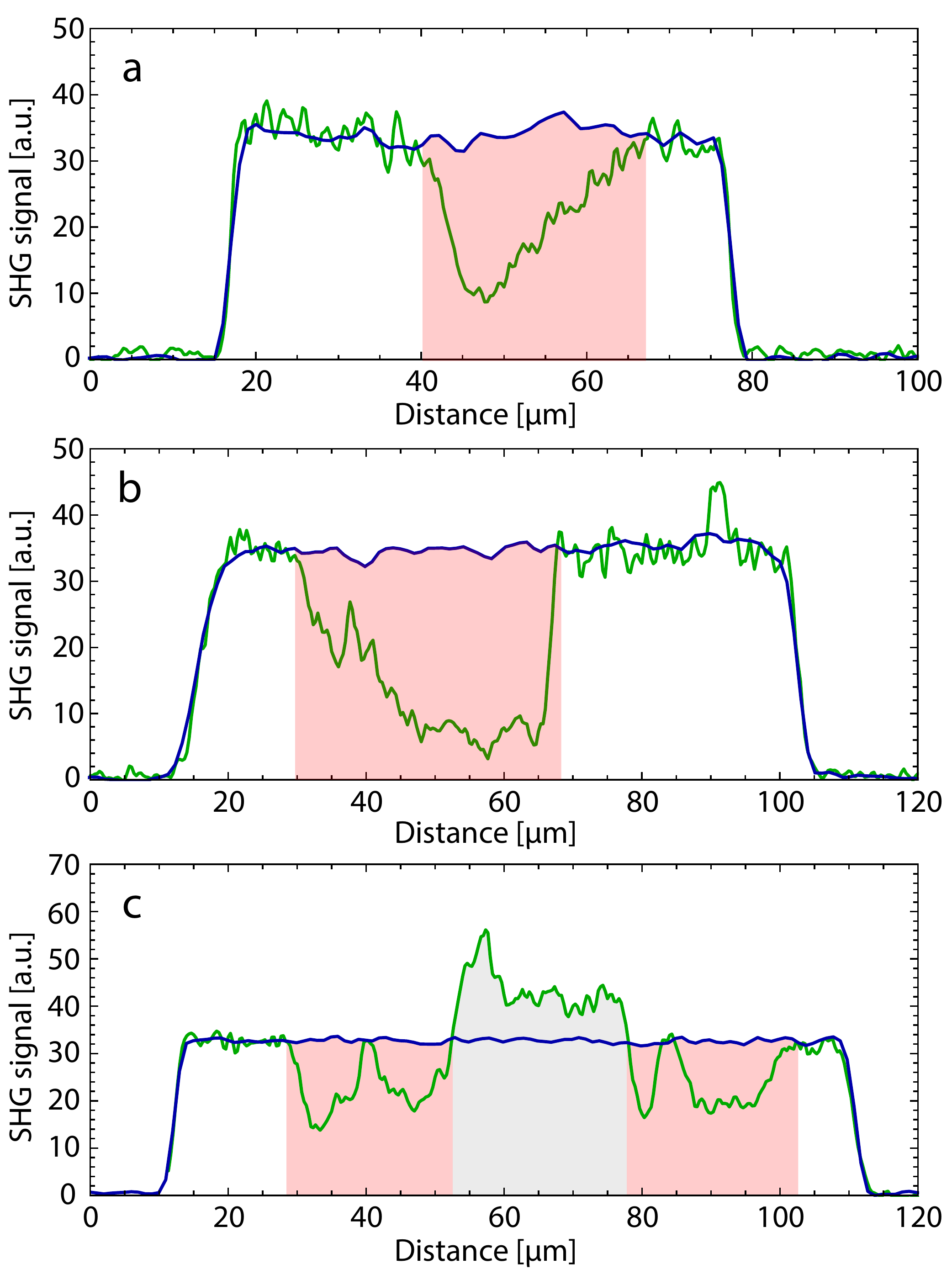}
\caption{\label{fig:cut} Vertical (a) and horizontal (b, c) cross-sections of the SHG signal depicted in Fig.~\ref{fig:Sample}. Blue and green lines correspond to the SHG response before (Fig.~\ref{fig:2x2}\,b) and after (Fig.~\ref{fig:2x2}\,d) the incommensurate-to-commensurate phase transition, respectively. Light red areas highlight the reduction of the SHG signal due to a presence of the aligned graphene. The enhanced response at gray area indicates the commensurate multilayer graphene. The value of a signal is given in arbitrary units.}
\end{figure}

To modify the alignment of the sample, we exploit the finding of the work~\cite{woods2016macroscopic}, where thermal annealing was used to activate the incommensurate-to-commensurate phase transition. For this aim, we induce a local laser heating of the graphene/hBN heterostructure via a long irradiation of the sample. This was done in such a way that the system was under the irradiation for tens of seconds. The estimated energy of the irradiation is $\sim$0.06 J/cm$^2$ (the pulse energy is 1.88 nJ, the spot size area is  $\sim$ 1 $\mu$m$^2$). Heated by a laser, the structure relaxed from the incommensurate state to a more favorable commensurate one. After that, a Raman spectroscopy measurement was performed to confirm the phase transition. It has been observed that single-layer graphene flakes became aligned with the underlying hBN forming moir\'e patterns with the periodicity of 12.5 and 14.0 nm. The alignment of a multilayer graphene has not been determined by Raman spectroscopy. For more details see the SI~\cite{SI}. 

Now, we repeat TPL and SHG measurements. 
Corresponding results are shown in Fig.~\ref{fig:2x2}\,c and d, respectively. 
As expected, the TPL signal remains unchanged after the phase transition. This follows from the fact that the electronic band structure of the system at a characteristic energy of nonlinear optical excitations does not change under small rotations~\cite{PhysRevB.89.201404}.
On contrary, the SHG response is drastically modified after graphene becomes aligned with the hBN substrate.
Indeed, now the SHG intensity picture explicitly shows the position of graphene flakes.  
According to a theoretical prediction~\cite{PhysRevB.99.165432}, this is a clear signature of the inversion symmetry breaking that occurs due to the interaction of the aligned graphene with the underlying hBN layer. 
As a consequence, electrons in the commensurate phase of the graphene/hBN heterostructure gain a nonzero average mass~\cite{PhysRevB.87.245408, PhysRevB.84.195414, PhysRevLett.115.186801, Gorbachev448}, which is precisely captured by the SHG experiment. 
Remarkably, we find that the SHG response from the area of the multilayer graphene has also been modified indicating the phase transition towards the commensurate phase. 

A precise comparison of the normalized intensity of the SHG for different alignments is shown in Fig.~\ref{fig:cut}. Results are obtained for cross-sections along vertical (a) and horizontal (b and c) directions depicted by orange lines in Fig.~\ref{fig:Sample}. Blue and green lines in Fig.~\ref{fig:cut} indicate the SHG signal in the incommensurate and commensurate phases, respectively. 
Remarkably, we observe a completely different change of the SHG response from single- and multilayer graphene/hBN heterostructures after the phase transition.
Indeed, the SHG is suppressed at light red ares that correspond to single-layer graphene flakes.
On contrary, the SHG from the aligned multilayer graphene depicted by gray color is enhanced.   

It is worth noting that the electric-dipole polarization of the hBN surface has only the imaginary component, because the considered frequency of the light is below the band gap of this material~\cite{PhysRevB.99.165432}. This leads to a phase shift of the corresponding SHG signal with respect to the second-order optical response from the graphene/hBN interface, for which the electric-dipole polarization can have both, real and imaginary components~\cite{PhysRevB.99.165432}. Therefore, the interference between commensurate graphene and underlying hBN can be either destructive or constructive, depending on a particular structure of the system. This fact makes the SHG very appealing for investigation of the inversion symmetry breaking in layered heterostructures, because already {\it the change} of the SHG clearly indicates a structural phase transition. In this context, the Raman technique is a less direct method,
since it requires additional data processing steps, such as the calculation of the broadening of the 2D peak in the spectrum~\cite{eckmann2013raman}. 

For additional confirmation of obtained results, we have performed the TPL and SHG measurements of another aligned single-layer graphene encapsulated between two hBN flakes. In this case, we also observe a strong modification of the SHG signal associated with the presence of the aligned graphene flake between hBN layers. For more details of this experiment please see the SI~\cite{SI}.

In conclusion, we have observed that structural changes in graphene/hBN heterostructures can be explicitly captured using the optical SHG technique.
We realized that the incommensurate-to-commensurate phase transition in considered systems can be activated by a local laser heating of the sample. We have found that the SHG method is able to detect the alignment not only of the single-, but also of the multilayer graphene flake disposed on the hBN surface, contrary to Raman spectroscopy. In addition to transport measurements, our nonlinear optical study confirmed that the average mass of electrons in graphene is zero in the incommensurate phase, and is nonzero for the commensurate structure. Our results suggest that the proposed method for detection of the alignment is more direct and, thus, has an advantage over a standard Raman technique.

\section*{Methods}
\label{sec:Methods}

Samples for these measurements were  fabricated by dry-peel methodology described elsewhere~\cite{woods2014commensurate, kretinin2014}. Control of the alignment is achieved by positioning long, straight edges of the crystals (which tend to be one crystallographic axis, zig-zag or arm-chair) parallel to each other. We use single- and multilayer graphene flakes placed on top of thick hBN ($\sim150$\,nm).

Raman spectroscopy measurements were performed using a Horiba Raman spectrometer (grating 1200 GPI) operating with an incident laser at a wavelength of 532\,nm and $\sim0.5$\,mW power. A confocal microscope was used to focus on the sample through a $\times100$ objective. For more details please see the SI~\cite{SI}.

For nonlinear characterization of the samples WITec alpha300 S confocal microscope was used in reflection geometry. Samples were irradiated by Ti:sapphire oscillator at 800\,nm and $\sim100$\,fs pulse width. Typical laser power was $\sim220$\,mW before the microscope where about 70\% of power reached and were focused on a sample with $\times20$ Zeiss objective. Detected  nonlinear response was separated from fundamental wavelength by use of two types of filters. SCHOTT BG39 filter (390-650\,nm transmission) was used for experiments when SHG/TPL combined response was detected. Thorlabs FB400-40 filter was used in experiments where only SHG signal was of interest.

\begin{acknowledgments}
Authors thank Chris Berkhout for technical support. Authors also thank Clement Dutreix and Vladimir Kukushkin for fruitful discussions and useful comments. The work of E.A.S. was supported by the Russian Science Foundation, Grant No. 17-72-20041. K.S.N acknowledges support from EU Graphene Flagship Program (contract CNECTICT-604391), European Research Council Synergy Grant Hetero2D, the Royal Society, EPSRC grant EP/N010345/1. The work of M.I.K. was supported by European Research Council via Synergy Grant 854843 - FASTCORR.
\end{acknowledgments}

\bibliography{Ref}

\appendix
\clearpage
\onecolumngrid

\begin{center}
\Large{Supplemental Material for \\
``Nonlinear optical study of commensurability effects in graphene-hBN heterostructures''}\\[1cm]
\end{center}

\twocolumngrid

\section*{Raman characterisation of the sample}

\begin{figure}[t!]
\includegraphics[width=0.7\linewidth] {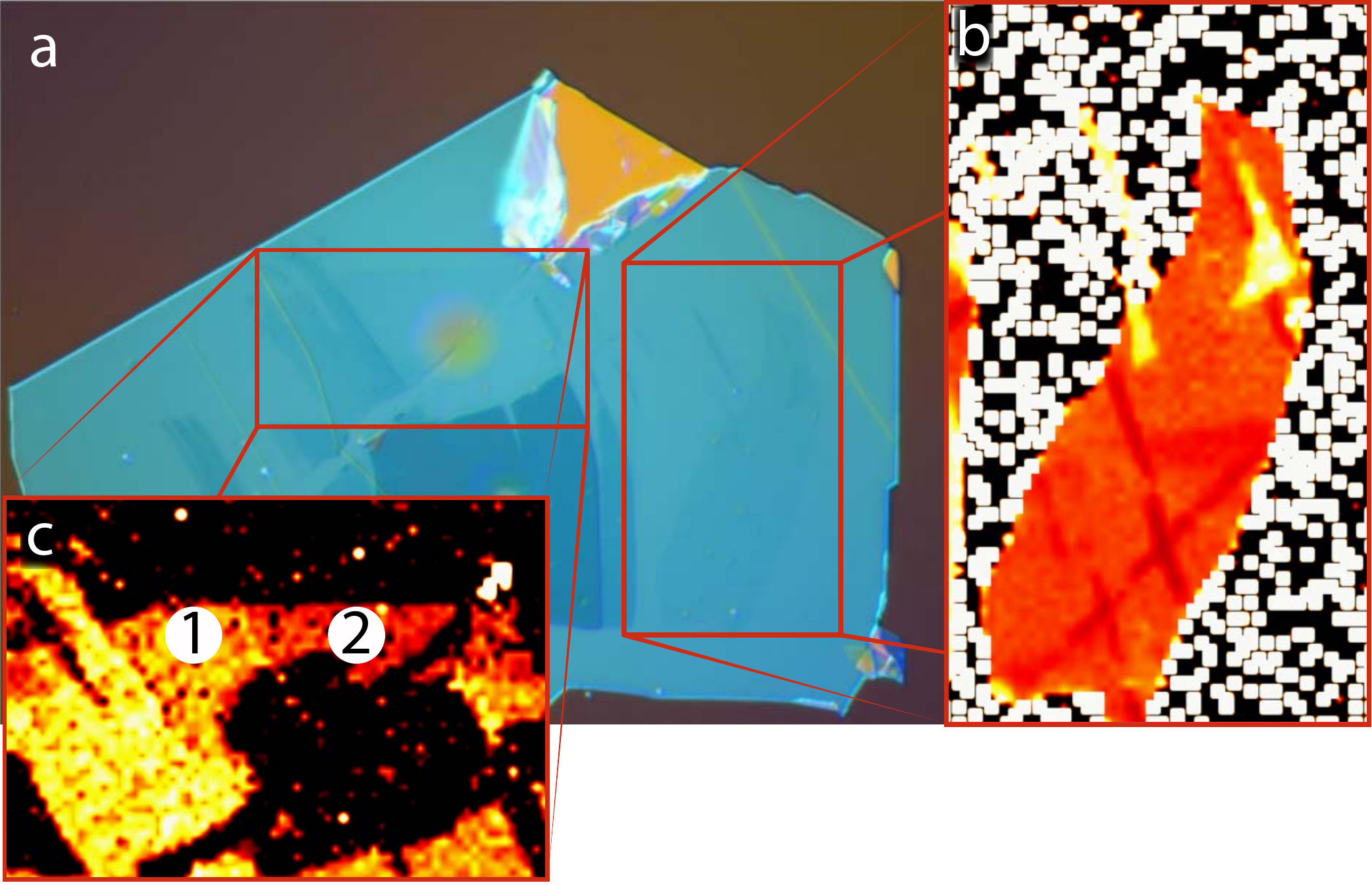}
\caption{Optical picture of the sample (a) and Raman spectroscopy for single-layer graphene flakes (b and c) after the phase transition. Labels (1) and (2) on the panel c indicate regions with different moir\'e periodicity specified in the text.}
\label{fig:periodicity}
\end{figure}

We confirm the presence or absence of a commensurate phase at the interface between the crystals by Raman spectroscopy. This is because the commensurate phase for graphene on hBN is characterised by the appearance of a strain distribution when the crystals near to alignment~\cite{woods2014commensurate}. The Raman spectrum of graphene is sensitive to slight changes in uniaxial/biaxial strain. In particular, the 2D-peak responds to the commensurate phases’ strain distribution by broadening~\cite{eckmann2013raman}.
Here, we observe that the full-width half-maximum of the 2D-peak (FWHM(2D)) before our experiments is 23~cm$^{-1}$, which is consistent with an unaligned flake ($>1.5^{\circ}$ misalignment)~\cite{eckmann2013raman}, or graphene on a rough substrate (SiO$_2$, polymers etc.)~\cite{ferrari2013raman}. This indicates that the graphene and hBN crystals cannot be in the commensurate phase.

After the phase transition, the FWHM(2D) shows significant change. 
The width of the peak has broadened to $\sim36$\,cm$^{-1}$, which is consistent with the most aligned case of graphene on hBN ($0.3^{\circ}$ alignment, moir\'e period is $L=13.5$\,nm). Corresponding results are shown in Fig.~\ref{fig:periodicity}. Here, insets b and c depict two single-layer graphene flakes. The moir\'e period obtained for the flake b is $L=12.5$\,nm (FWHM(2D) is 33\,cm$^{-1}$). Moir\'e periods for areas (1) and (2) of the flake c are $L=14.0$\,nm (FWHM(2D) is 40\,cm$^{-1}$) and $L=12.5$~nm (FWHM(2D) is 33\,cm$^{-1}$), respectively. This result demonstrates unambiguously that the graphene is now in a commensurate phase with the hBN crystal. 

\section*{Encapsulated graphene}

\begin{figure}[t!]
\includegraphics[width=0.7\linewidth]{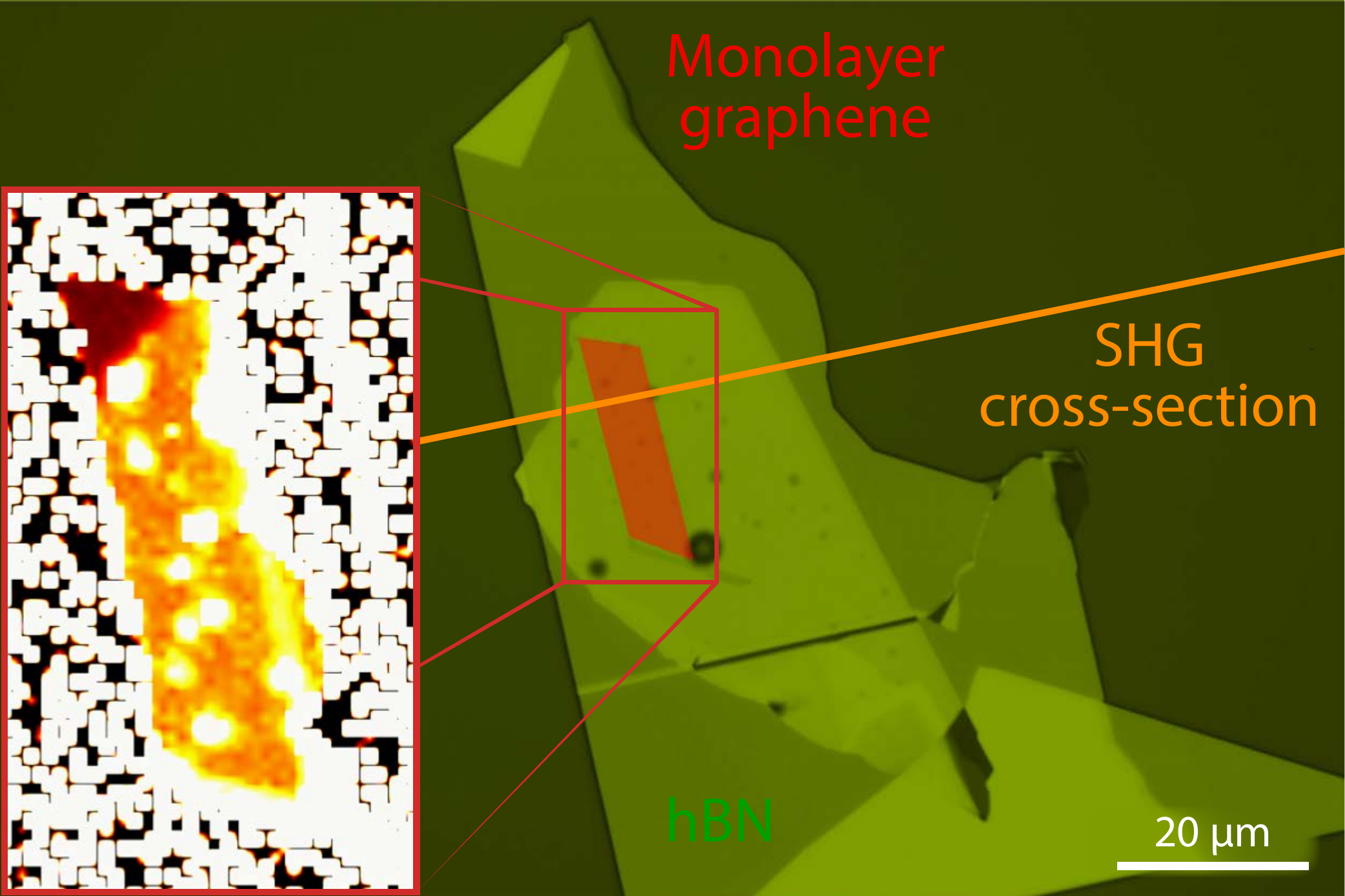}
\caption{Optical picture of the encapsulated and aligned graphene sample. A single-layer graphene flake is shown in red. The substrate hBN crystal is outlined in light green. Orange line indicates the cross sections shown in Fig.~\ref{fig:enc2}\,c. Scale bar is 20\,um.
The inset is a Raman spectroscopy result.}
\label{fig:enc}
\end{figure}

For additional confirmation of our findings, we repeat the nonlinear optical study on another sample.
The optical picture of the sample is shown in Fig.~\ref{fig:enc}.
Here, a red area highlights an aligned monolayer graphene encapsulated between two hBN layers depicted by a light green color.
The alignment of the graphene is confirmed by Raman spectroscopy. The moir\'e period is found to be $L=13.5$\,nm (FWHM(2D) is 37\,cm$^{-1}$). 

\begin{figure}[b!]
\includegraphics[width=0.7\linewidth]{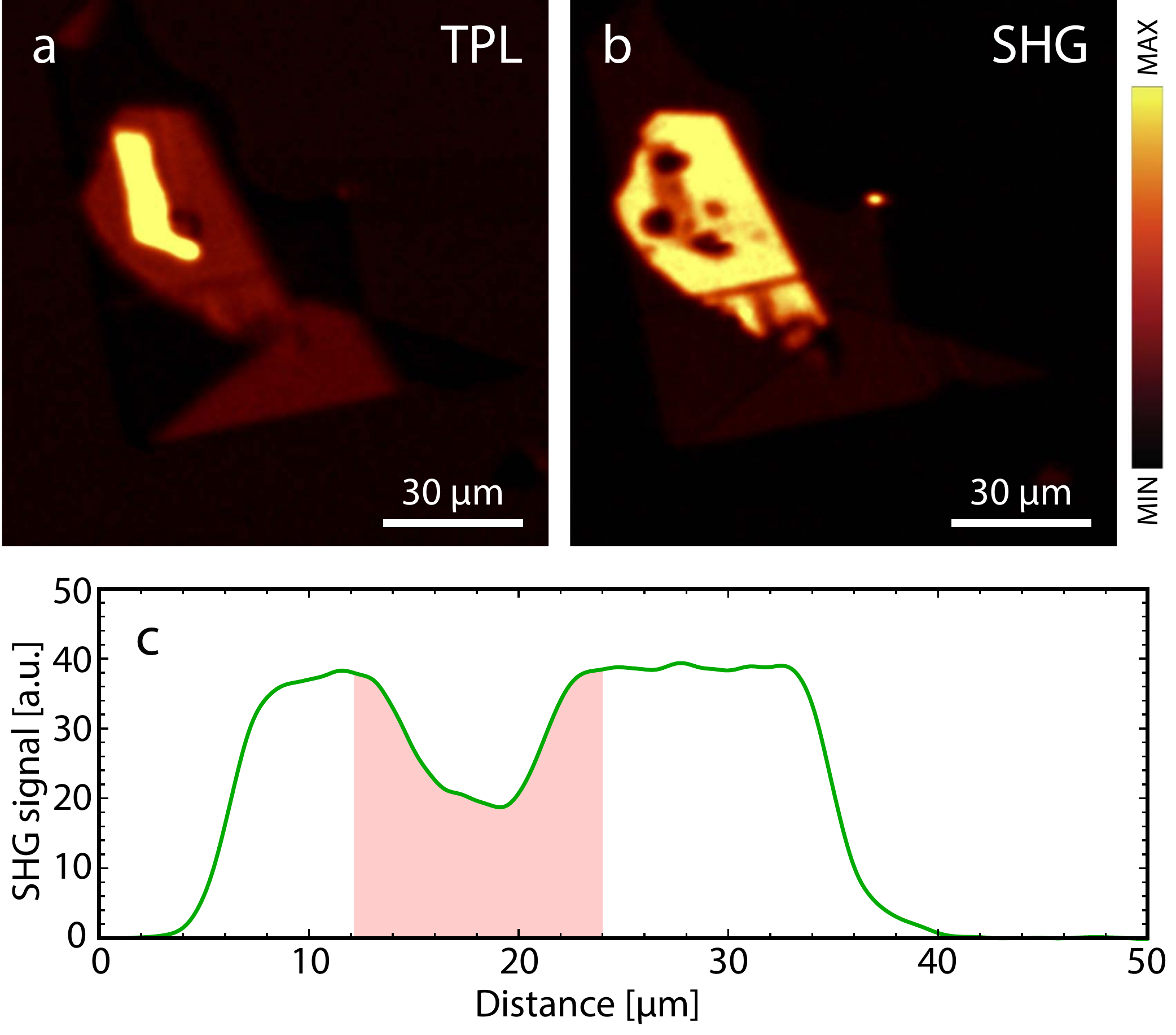}
\caption{\label{fig:enc2}
TPL (a) and SHG (b) signals from the aligned and encapsulated graphene/hBN heterostructure. Color bar depicts the intensity of the nonlinear response. A lighter color indicates a larger value of a signal. Panel (c) shows a cross-sections of the SHG signal depicted in Fig.~\ref{fig:enc}. Light red area highlights the reduction of the SHG signal due to a presence of the aligned graphene. The value of a signal is given in arbitrary units.
}
\end{figure}

The TPL and SHG intensity pictures are shown in Fig.~\ref{fig:enc2}\,a and b, respectively. Here, both, the TPL and SHG responses clearly shows the presence of a graphene flake. The change of the SHG signal from graphene with respect to the hBN environment confirms that the monolayer graphene flake is in the commensurate phase. The strong suppression of the SHG can be explicitly seen in Fig.~\ref{fig:enc2}\,c that shows the cross-section of the SHG signal depicted by the orange line in Fig.~\ref{fig:enc}. Here, the light red area corresponds to the position of the aligned single-layer graphene flake.

\end{document}